\documentclass[12pt]{aastex}
\usepackage{natbib}

\shorttitle{Russell Lecture}
\shortauthors{Donald Lynden-Bell}
\begin{document}
\title{Russell Lecture: Dark Star Formation and Cooling Instability}
\author{{D. Lynden-Bell}\altaffilmark{1}}
\affil{Institute of Astronomy, The Observatories, Madingley Road,
Cambridge, CB3 0HA, UK}
\altaffiltext{1}{and Clare College, Cambridge, UK}
\email{dlb@ast.cam.ac.uk}

\and

\author{{C. A. Tout}\altaffilmark{2}}
\affil{Institute of Astronomy, The Observatories, Madingley Road,
Cambridge, CB3 0HA, UK}
\altaffiltext{2}{and Churchill College, Cambridge, UK}
\email{cat@ast.cam.ac.uk}
\begin{abstract}
Optically thin cooling gas at most temperatures above 30K will make
condensations by pressure pushing material into cool dense regions.
This works without gravity.  Cooling condensations will flatten and
become planar/similarity solutions.  Most star formation may start
from cooling condensations - with gravity only important in the later
stages.  The idea that some of the dark matter could be pristine white
dwarfs that condensed slowly on to planetary sized seeds without
firing nuclear reactions is found lacking.  However, recent
observations indicate fifty times more halo white dwarfs than have
been previously acknowledged; enough to make the halo fraction
observed as MACHOS.

A cosmological census shows that only 1\% of the mass of the Universe
is of known constitution.
\end{abstract}
\section{Introduction - Russell in England 1902--1905}
After some time spent in attending lectures, Russell joined
A. R. Hinks, a skilled observer, with whom he learned the techniques
of accurate parallax determination.  They used the Sheepshanks
telescope, which cost\footnotemark[1]\footnotetext[1]{To turn these
prices into current dollars, multiply by about 100} \pounds3,183 in
1898.  It was still at the Observatories for my early years in
Cambridge, but was demolished in 1959.  W. M. Smart, who used the
telescope extensively, wryly remarked
\begin{quote}
``It was a telescope of unusual design combining in a unique way the
chief disadvantages of both the refractor and the reflector!''
\end{quote}
In 1903 Hinks was promoted to Chief Assistant at a salary of
\pounds220\footnotemark[1] per annum and a house provided.  Russell's
Carnegie (post doctoral) grant was \pounds200\footnotemark[1] per
annum.

It is a tribute to the persistence of Hinks and Russell that they
eventually determined 50 parallaxes, many of which stand up well
to comparison with modern determinations.  A not atypical example is
the parallax of the well known high velocity subdwarf Groombridge 1830
for which Russell gives $0.100 \pm 0.029^{\prime\prime}$ as compared to
a modern value of $0.107^{\prime\prime}$.  Thus, many of the stars in
Russell's first HR diagram were put there on the basis of their
Cambridge parallaxes. 
\section{Census of the Universe}
The prediction that the angular wavelength of the first peak in the
angular power spectrum of the cosmic microwave background would decide
whether the Universe was closed or open was made only 14 years ago
\citep{efs86,bon87} [see also \citet{dor78}].  The first results of
TOCO, Boomerang and Maxima are already with us.  By contrast the
prediction that many large galaxies including our own, M31, M32, M81,
M82, M87, etc, contain giant-black-hole remnants of dead quasars in
their nuclei \citep{dlb69} took over 26 years before the first
definitive case was found \citep{miy95}.  Although there were
indicative results earlier \citep{you78,sar78,pen81,pou86,dre88,kor88}
and many convincing ones since (\citet{eck96, gen97, geb00}).  An
interesting account of twentieth century developments is given in
\citet{shi99} review, but it can be argued that the first prediction
was 211 years earlier \citep{mic1784}.

Astronomy is now running so fast that almost everything is as
ephemeral as a news report, nevertheless each entry in Table 1 is
attested by at least two independent lines of evidence.  For example,
the cold dark matter fraction comes both from estimates of the baryon
fraction in large $X$-ray emitting clusters of galaxies
\citep{whi93} and from estimates of the relative height of the peaks
in the Maxima results.  Likewise the unknown expansion energy estimate
comes both from the difference between the closure density (TOCO,
Boomerang and Maxima) and the total matter density of Maxima, {\bf
and} from the rate of acceleration of the Universe measured using
distant supernovae.  

However, if in a thought experiment we dream of reversing the Hubble
expansion -- but nevertheless leaving time running forward -- then the
black body radiation from the sky would get hotter and hotter.  The
white dwarfs, neutron stars and black holes would all survive to
temperatures of more than $10^9$K and there is then no time to destroy them
before the big crunch.
\begin{table}
\begin{center}
\label{tbl-1}
\begin{tabular}{llll}
\tableline\tableline
TIME & \multicolumn{2}{c}{FRACTION $\Omega = \rho/\rho_c$} & $\Omega$ \\
\tableline
$H^{-1}_o =10^{10}h^{-1}$yrs & \multicolumn{2}{c}{$\rho_c = 3H^2_o/(8\pi
G)$} & $h = 0.6$\\
10sec & Baryons (Atoms) & $1.9h^{-2}\%$ & $5\%$ \\
		&	& $\pm 0.1$ & \\
$1/3$ Myr & Baryons (Atoms) &  $3.2h^{-2}\%$ & $9\%$ \\
		&	& $\pm 0.6$ & \\
& Dark Matter + Baryons	&  $15h^{-2}\%$ & $40\%$ \\
		&	& $\pm 5$ & \\
$10^{10}$ yr & Radiation & & $0.007\%$\\
		& Neutrinos & & $1\%$\ ?\\
		& Unknown Expansion & & \\
		& Energy $\Omega_\Lambda$ (Supernovae) &&  $\sim
60\%$\\
$1/3 $ Myr & Closure & $1.03 \pm 0.1$ & $100\%$ \\
\tableline
\end{tabular}
\caption{Universal constituents and effective times of measurements.}
\end{center}
\end{table}
This thought experiment shows us that any compact bodies which might
already be present at the time of Helium creation 10 seconds after the
Big Bang would survive to the present day.  If so, their baryons would
not be counted in the first two entries but would contribute to the
collisionless matter of the third entry.  The two entries for baryons
disagree but the second estimate is based on the results of Boomerang
and Maxima which don't agree well either.  Few believe the number of
baryons goes up between 10 sec and 1/3 Myr and the first entry is
considerably more secure.  However, even that gives us a {\bf missing
baryon problem} because the numbers of baryons in stars, galaxies and
intra-cluster gas is much less than that given by the first entry.
Much may lie as yet undetected as $5.10^5$ K gas in small groups of
galaxies like ours.  Detection of the dispersion measure to pulsars in
the globular clusters of the Fornax dwarf would test this.  

The wonder of astronomy lies as much in what is unknown as in the
beauty of the 1\% that is known.  However, when discussing these
unknowns, one should heed the wisdom of the past, so when lecturing
about cold dark matter, I always quote
\begin{quote}
``If a thousand men believe a foolish thing it is still a foolish
thing.''
\end{quote}
A new way of looking for invisible heavy halo objects via
gravitational lensing events towards the Magellanic Clouds was first
proposed by \citet{pet81}.  Independent calculations were made
 by \citet{pac86}, and it was he who stimulated \citet{alc96} who first
convinced the world that it was a practical possibility and led the
Macho team to find such events \citep{alc00}.

The Macho team found over twenty events toward the LMC which could
account for between 10\% and 20\% of the Galaxy's dark halo, although
almost half of the events might be self-lensing within the LMC.
Whereas they found too few events to account for their halo model,
both they and the Eros and Ogle teams found more than the expected
number of lensing events towards the Galaxy's bulge.  It would be not
unnatural to imagine that the models of the Galaxy's mass distribution
were seriously wrong with too little in the disc and too much in the
halo, but that is not how the experts put it.

While pondering the nature of dark matter and the Macho results on Feb
15th 2000, I realised that there was a possibility that I had never
heard discussed,
\begin{quote}
``Could  white dwarfs be made directly from
primordial material without ever forming stars or burning hydrogen?
Could such pristine white dwarfs be the halo objects found by Macho?''
\end{quote}
As we shall see, such questions had occurred earlier to \citet{sal92}
and \citet{han99}.  
\section{Pristine and Halo White Dwarfs}
Discussions of star formation often start with a large body
contracting from a diffuse cloud.  The equilibrium radius of such a
body is determined by the Virial theorem.  
\begin{displaymath}
{\cal T} = - {{1}\over{2}}{\cal V}= - E \ ,
\end{displaymath}
where $\cal T$ the kinetic energy is ${{\scriptstyle{3}\over{2}}}N
k\bar{T}$ and $\cal V$ the potential energy is $-GM^2/(2\bar{R})$.
The radiative loss from the surface causes the energy to get more
negative so the radius $\bar{R}$ decreases and the internal
temperature $\bar{T}$ increases until the nuclear reactions fire and the body
becomes a star.

The above conclusion can be circumvented if the kinetic energy is not
due to the star's temperature but is determined by the zero point
energy of confined electrons required by the uncertainty principle.
This occurs in solids and liquids.  

If I were to freeze the A.A.S. President to absolute zero, she might
become glassy eyed and she would shrink by an inch or so, but not
substantially.  This is because the electrons in her atoms owe their
kinetic energy neither to their temperature nor to their angular
momentum (since there is none in e.g., the ground state of hydrogen)
but to their zero point degeneracy energy.  They are of course held
together by electricity.  

Asteroids, planets and white dwarfs all owe their sizes to this zero
point energy of electrons.  In fact due to the work by \citet{fow26},
\citet{sto29}, \citet{and29} and \citet{cha31} and by
\citet{sal67} in the 1960's, the mass radius relationship for such
`cold' bodies is one of the best understood parts of astrophysics.

In outline there are three terms: 
\begin{itemize}
\item The zero point energy of the electrons which may or may not be
relativistic. 
\item Their electrical binding to their associated atomic nuclei.
\item The gravitational potential energy.
\end{itemize}
In small bodies such as asteroids, the main balance is between the
zero point energy and the electrical attraction of the nuclei.
Gravity merely serves to keep the parts together so as the bodies are
piled together the mean density remains much the same.  In spite of
the increased pressure at the centre this equal density law still
holds up to masses close to Saturn's.  About there the gravitational
binding starts to influence the total binding energy which is still
mainly electrical.  As a result the mean density starts to increase.
At a mass a little over the mass of Jupiter gravitational and
electrical binding become equal and the weight of each extra mass
added causes a contraction as great as its volume.  Beyond that point the
bodies get smaller as the mass increases.  When the zero point energy
of each electron is substantially less than its rest mass, we get the
brown dwarf--white dwarf sequence with $R \propto M^{-1/3}$.  Once the
electrons become relativistic the pressure generated becomes as soft
as gravity and the radius decreases very rapidly as the Chandrasekhar
mass is approached  \citep{and29, sto30, cha31}.  It is possible to get
a single mathematical formula for the whole mass-radius relationship
from asteroids to Chandrasekhar's limit \citep{dlb01}.
\begin{displaymath}
\left ({{4}\over{3}}\pi \rho_0 \right )^{1/3}R =
{{M^{1/3}}\over{1+(M/M_p)^{2/3}}}I \ ,
\end{displaymath}
where $M_p$ is the mass of the planet of maximum radius and $\rho_0$ is
the density of the solid at zero pressure.  $I$ is unity except
close to the Chandrasekhar limit $M_{ch}$ where
$I=J/[1+(3/4)(1-J^{1/2})]$ with $J=\sqrt{1-(M/M_{ch})^{4/3}}$.

The mass of the planet of maximum radius is set by the equality of
gravitational and electrical attractions.
\begin{displaymath}
M_p \approx \left ({{e^2}\over{Gm^2_H}}\right )^{3/2} m_H \ .
\end{displaymath}
The Chandrasekhar mass is approximately $(\hbar c/e^2)^{3/2}$ times
greater than this, showing that the fine structure constant plays a
vital role.

In the theory of gravitational fragmentation the mass of the
non-fragmenting fragment is given by \citep{low76}
\begin{displaymath}
M= \left [{{3^5 \cdot 5^4}\over{2^8\pi^5}} \left
({{\hbar c}\over{e^2}}\right )\ \left ({{hc}\over{Gm^2_H}}\right )^{10} \left
({{m_e}\over{m_H}}\right )^2 \right ]^{1/7} m_H \simeq M_\odot/60 \ ,
\end{displaymath}
which may determine the minimum mass of bodies so formed, but as yet
there is little evidence that nature pays too much attention!  See in
particular the fascinating $\sigma$ Orionis Cluster \citep{zap00}.

Figure 1 shows two mass-radius relationships; that on the left is
appropriate for rock or for white-dwarfs with $(\mu =2)$ two baryon
masses per electron.  That on the right $(\mu = 8/7)$ is appropriate
for a pristine mixture of 75\% Hydrogen and 25\% Helium by mass, i.e., for
Hydrogen planets, brown dwarfs and - if such exist - pristine white
dwarfs.  The latter have radii $R\propto \mu^{-5/3}M^{-1}$ that is 2.5
times larger than equal mass normal Helium, Carbon or Oxygen white
dwarfs.  At the same surface temperature we would expect them to be
two magnitudes brighter.  Since they have twelve times as many nuclei
as Carbon white dwarfs their specific heats will be greater and they
may not cool faster in spite of this extra emission.  

When first interested in pristine white dwarfs I consulted early works
by \citet{par46} and by \citet{egg65} which showed two sequences
separated by nearly two magnitudes.  It was rather ironical to find
that after Eggen consulted a theorist his work on the Hyades
white dwarfs restricted itself to candidates within one magnitude of the
expected sequence and would therefore have missed a second sequence.
The idea was that in low-density regions such as cooling flows, galaxy
haloes and dwarf spheroidals, stars form very gradually by accretion
on to planetary sized seeds.  If the growth was {\bf so} slow that the
accretion energy was radiated before the material was buried then the
body would grow in mass resting on its zero point energy and never
getting hot enough to fire its nuclear fuel.

When L-B raised the question of making pristine white dwarfs with
C. A. Tout, his immediate reaction was that they would blow up like novae,
but a day later he had found that novae only explode when the
temperature of the degenerate hydrogen layer exceeds $10^6$K.  If the
accretion rate were so small that the temperature never reached a
million then pristine white dwarfs might be made.

To see if any had been observed we next consulted Mike
Irwin.  He said ``it is possible that we have just what you are looking
for'' and sent us to read the growing literature on halo white dwarfs.

When \citet{iba99} repeated the Hubble deep field three years later
they found five objects at $29^{\rm th}$ magnitude which appeared to move.  One
may be a distant supernova in the side of a pre-existing image, two of
the other images have unexpected asymmetries, but two looked like
single stars\footnote{Since this lecture was given, a
third epoch HST frame has shown that neither of these remaining
objects has a uniform motion, but the objects found by Schmidt surveys
are real and more candidates have since been found (see \citet{opp01}
and the criticisms of \citet{rei01}).} that move 20 milliarc
sec/yr.

If such a star were brought 100 times closer it would be at $19^{{\rm th}}$
magnitude and move 2 arc sec/yr.  Thus Irwin planned a wide--field
Schmidt Survey to find such rapidly moving objects from existing plate
material with plates taken less than ten years apart so that the faint
stars would not move so far that they were `lost' among others.

After searching over 700 square degrees \citet{ibayr} found 18
possible objects of which the brightest two were investigated
spectroscopically.  Both showed the strong infra-red blanketing
predicted as the hallmark of molecular hydrogen formation in cool
white dwarf atmospheres (\citet{mou78, han98}).  Both had motions
appropriate for halo stars `left behind' by the rotation of the
Galaxy.  Earlier \citet{hod00} had found another white dwarf showing
the strong predicted infra-red blanketing.  To be so fast moving and
so dim the objects cannot be main-sequence stars without leaving the
Galaxy.  Making them old white dwarfs means they are quite close,
about 30 parsecs.  This means such objects are common; 50 times more
common than earlier estimates of halo white dwarfs, but not all
estimates ({\citet{cha96, cha99}) showed that if the halo mass
function peaked a little above $1 M_\odot$ then there could be many
cool old white dwarfs.  At such a density they would account for up to
10\% of the dark halo.  Others have objected that the population of
heavier stars born with them would have produced more metals than are
observed, but that would be avoided if these were pristine white
dwarfs.  At this juncture it seemed that the dark curtain over the
nature of dark matter had let through a chink of light.  L-B started
to study how stars might form at low densities in cooling flows but,
in the meantime, Tout was in real difficulties.  Pristine white dwarfs
over about $0.2 M_\odot$ always fired their hydrogen unless their
accretion rates were only 1 $M_\odot/10^{13}$ years -- too slow to be
made in this Universe.  Furthermore this agreed with the estimates
made by \citet{sal92} on the basis of the accretion models of
\citet{len92}.  Just as nuclear physicists predict an island of almost
stable ultra-heavy nuclei that are very hard to make, so pristine
white dwarfs between $0.2M_\odot$ and the cold pycnonuclear reaction
limit of $1.1M_\odot$ are a theoretical possibility, but no-one knows
how to make them!

This makes yet more interesting the origin of the large number of halo
white dwarfs now being found.  (See also \citet{har01}).

\section{Cooling Instability}
Although the investigations of sections 4-7 were stimulated by the
consideration of pristine white dwarfs, they are independent and
develop the non-linear theory of cooling instabilities which are
probably important precursors of galaxy formation, star-cluster
formation and normal star formation.

As stars are held together by gravity it is natural to believe that
gravity is the primary driver in their condensation, but this may be
wrong.  At a density of $10^{-22}{\rm g\ cm}^{-3}$ a solar mass of
gas has an escape velocity of only 0.1 km/s whereas an ionised medium
of that density will have a sound speed of about 10 km/s.  At low
densities pressure can be much more important than gravity and the
pressure is often dictated by the cooling as the gas radiates to
infinity \citep{fab94}.  

Anyone who looks at photographs knows that the idea of a uniform
interstellar medium is a theoretical abstraction far from truth.  With
such variations from place to place the cooling rate per gram will
vary.  This will induce pressure differences that will push the
interstellar gas to the low pressure regions.  The cooling is commonly
such that regions of lower pressure have higher density.  Thus
condensations can form due to pressure alone, quite independently of
any gravity.  This is how condensation begins, the gravity comes in
later.

We shall consider the simplest possible case of an optically thin gas
at pressure $p$ cooling by radiation to infinity.  Suppose there are
regions of different densities with anti-correlated temperatures so
that dynamical equilibrium is maintained with the pressure spatially
uniform.  The second law of thermodynamics gives the entropy decrease
consequent on the cooling $\rho^2\Lambda(T)$ per unit volume; so for unit
mass
\begin{equation}
T\ Ds/Dt = -\rho \Lambda (T) \ .
\label{eq:4.1}
\end{equation} 
For a perfect gas, $s=km^{-1} \ln (T^{3/2}/\rho) + {\rm constant} =
km^{-1}\ln (T^{5/2}/p) + {\rm constant}$ where $m$ is the molecular
mass so for two regions of gas at a common pressure $p=m^{-1}kT\rho$, 
\begin{equation}
(5/2) k^2m^{-2} p^{-1} D/Dt \ln (T_1/T_2) =
T^{-2}_2\Lambda(T_2)-T^{-2}_1\Lambda(T_1)\ .
\label{eq:4.2}
\end{equation}
Hence as the gas cools the temperature ratio will increase whenever
$T^{-2}\Lambda(T)$ is a decreasing function of temperature.  This is
the cooling instability criterion discovered by \citet{fal85}.  As
they pointed out it differs from the $T^{-1}\Lambda(T)$ decreasing
criterion for thermal instability found in Field's masterly paper
\citep{fie65} and originally due to \citet{wey60}.  The reason for the
difference is that Weyman and Field studied small departures from
equilibrium conditions in which heating and cooling balanced in the
mean state while the cooling instability criterion holds when
everything is cooling on a timescale that is slow enough that the two
regions share a common pressure (which may vary with time).  

Figure 2, adapted from \citet{dal72}, shows that for $30{\rm K} <T<9000{\rm K}$ and $T>10^4{\rm
K}$ the whole range in which gas is ionised, $T^{-2}\Lambda(T)$
decreases. We deduce that in the absence of heating, cooling
instabilities will occur and cooling condensations will develop.
However, we have neglected two effects that work to suppress the
instability.  At small scales thermal conductivity will tend to
suppress temperature differences and thus depress the instability by
heating the denser regions.  But, once the instability is kick started
the thermal conductivity falls as the temperature falls so limiting
the effect.  More importantly \citet{bal89} have shown that, for
infinitesimal perturbations, there are no thermal instabilities when
the gas is in stratified equilibrium in a gravitational field.  How
can this be?  Surely if we make the gravity field weaker and weaker we
must arrive at the uniform medium instability criterion once the
gravitationally induced gradients are weak enough!  Here it becomes
necessary to contrast mathematically exact results for infinitesimal
perturbations with physical results for finite perturbations
\citep{luf95}.  An Astrophysicist might well consider a perturbation
with $\delta \rho/\rho$ of 1/10 or even 1/3 to be quite small, but a
Mathematician will consider only the limit as $\delta \rho/\rho$ tends to
zero so that all its gradients are small compared with those that he
has in his unperturbed state.  For weak gravity fields this is a
severe restriction on perturbation amplitudes especially at short
wavelengths.  We find in the next section that new density maxima
caused by the perturbation are all important.  In an infinitesimal
perturbation new density maxima only occur in a uniform medium.
However small the gradient in the unperturbed state, infinitesimal
perturbations make no new density maxima.  For systems with a gravity
field our results will only apply to finite amplitude perturbations
which make new maxima.
\section{Non-Linear Cooling Condensations}
We take the simplest possible case of no external gravity, no
self-gravity, no thermal conductivity and slow cooling so that the
pressure is spatially constant.  We consider this not because it is
the most realistic case but because it gives insight and {\bf insight}
rather than complicated exactness is what science is about
\citep{edd26}.  Once the insight is obtained more complications are
more readily understood.  For a plasma cooling by free--free emission
the cooling rate per gram can be written $\tilde{K}\rho c_s$ where $c_s$
the sound speed is proportional to $T^{1/2}$.  The coefficient
$\tilde{K}$ has the dimensions $[M^{-1}L^4T^{-2}]$ which is one power
of $L$ more than the dimensions of $G$.  As an aside it is interesting
to ask how long this length $\tilde{K}/G$ is, since both free--free
emission and gravity are inevitably involved in galaxy formation.  In
fact in fundamental dimensionless constants and the classical electron
radius, 
\begin{displaymath}
\tilde{K}/G = {{16}\over{3}} \left ({{2\pi}\over{3}}\
{{m_e}\over{m_p}}\right )^{1/2}\left ({{e^2}\over{\hbar c}}\right ) \
\left ({{e^2}\over{Gm_em_p}}\right ) {{e^2}\over{m_ec^2}} = 290 \ {\rm
kpc}\ .
\end{displaymath}
It is this large distance or rather 1/3.8 times it, that gives the 73
kpc radius from which galaxies fall together because their cooling
rate is faster than their collapse time.  We wish to discuss more
general cooling laws but, although it is not hard to be more general,
simplicity suggests a general power law $\propto \rho T^{2-\alpha}$ so the
above case corresponds to $\alpha = 3/2$.  With our general power law
equations (\ref{eq:4.1}) and (\ref{eq:4.2}) together with
the perfect gas law take the form 
\begin{equation}
5/2\ D/Dt \ln \left (p^{3/5}/\rho\right ) = -K \left (\rho/p^{3/5}\right
)^\alpha p^{1-2\alpha/5} \ ,
\label{eq:5.1}
\end{equation}
so
\begin{equation}
D/Dt \left (p^{3/5}/\rho\right )^\alpha = -(2/5)\alpha K p^{1-2\alpha/5}
\ .
\label{eq:5.2}
\end{equation}
Writing $\rho_0$ for the initial value of $\rho$ on the fluid element
considered and momentarily considering the special case with $p$
independent of time we find
\begin{displaymath}
(p^{3/5}/\rho)^\alpha = (p^{3/5}/\rho_0)^\alpha - (2/5)\alpha
Kp^{1-2\alpha/5}t
\end{displaymath}
and thence
\begin{equation}
\rho/\rho_0 = \left [1-(2/5) \alpha Kp^{1-\alpha}\rho^\alpha_0 t \right
]^{-1/\alpha}
\label{eq:5.3}
\end{equation}
However (\ref{eq:5.2}) may also be solved when
$p$ depends on time.  We introduce a new weighted time $\tau$ defined
by 
\begin{displaymath}
\tau = \int(p/p_0)^{1-2\alpha/5} dt \ ,
\end{displaymath}
where $p=p(t)$ and $p_0$ is its initial value.

Then in place of (\ref{eq:5.3}) we have the
general solution 
\begin{equation}
\rho/\rho_0 = (p/p_0)^{3/5} \left [1-(\tau/\tau_c)\right ]^{-1/\alpha} \ ,
\label{eq:5.4}
\end{equation}
where $\tau_c = 5/(2\alpha Kp^{1-\alpha}_0 \rho^\alpha_0)$ is the `time'
at which the density of our fluid element become formally infinite.
With the Fall-Rees criterion satisfied $\alpha>0$ so
this time is shorter when the initial density $\rho_0$ is greater.  Thus
nucleation occurs around maxima in the initial density distribution.
We rewrite (\ref{eq:5.4}) in terms of $\tau_m$ the
collapse time for the initial density maximum $\rho_m$.  Multiplying
(\ref{eq:5.4}) by $\rho_0/\rho_m$ we have
\begin{equation}
\rho/\rho_m = (p/p_0)^{3/5}\left [(\rho_0/\rho_m)^{-\alpha} -
\tau/\tau_m \right ]^{-1/\alpha} \ .
\label{eq:5.5}
\end{equation}
At time $\tau$ the current maximum density of the material whose
initial density was $\rho_m$ we call $\rho_c(\tau)$ or $\rho_c$ for
short.  For the maximum density we put $\rho_0 = \rho_m$ in
(\ref{eq:5.5}) and have
\begin{equation}
\rho_c/\rho_m = (p/p_0)^{3/5} \vert 1-\tau/\tau_m\vert^{-1/\alpha}
\label{eq:5.6}
\end{equation}
for $\tau <\tau_m$.  Of course for $\tau >\tau_m$ the maximum density
is infinite but we find it convenient to define $\rho_c$ as a
characteristic density given at all times by 
(\ref{eq:5.6})  even when $\tau > \tau_m$.  This
characteristic density is the maximum density at all times prior to
$\tau = \tau_m$ and increases.  After $\tau=\tau_m$ it becomes
smaller again and represents the density at a self-similar point in
the ensuing accretion flow.  Dividing (\ref{eq:5.5}) by  (\ref{eq:5.6})
we obtain
\begin{equation}
{{\rho}\over{\rho_c}} = \left [{{(\rho_0/\rho_m)^{-\alpha}-1}\over {\vert
1-\tau/\tau_m\vert }}\pm 1 \right ]^{-1/\alpha} \ ,
\label{eq:5.7}
\end{equation}
where the plus sign is to be taken for $\tau<\tau_m$, before the centre
collapses and the minus sign afterward.  As $\tau \rightarrow \tau_m$ the
denominator becomes small, so  (\ref{eq:5.7}) is then
very sensitive to small differences between $\rho_0$ and $\rho_m$.  This
means that the density profile near the collapse time $\tau_m$ depends
solely on the initial profile of $\rho_0$ close to $\rho_m$.  We now show
how this leads to similarity solutions.

Let $M(>\rho_0)$ be initially the total mass, centred on one maximum of
density $\rho_m$, which has at each point a density greater than some
chosen value $\rho_0$.  For a spherical density distribution around the
maximum we have 
\begin{displaymath}
dM=4\pi r^2 \rho_0 dr = (dM/d\rho_0)d\rho_0 \ .
\end{displaymath}
For a cylindrical distribution we replace $4\pi r^2 \rho_0 dr$ by $2\pi
R\rho_0 dR$ while for a planar distribution we would replace it by
$2\rho_0dz$ and measure mass per unit area.  Thus
\begin{equation}
\left.
\begin{array}{r}
(4/3)\pi r^3 \\
\pi R^2  \\
2\vert z \vert
\end{array}
\right \}
=\int^{\rho_m}_{\rho_0}(\rho^\prime_0)^{-1} (dM/d\rho^\prime_0)d\rho^\prime_0 =
\int^M_0 \left [\rho_0 (M^\prime)\right ]^{-1}dM^\prime \ ,
\label{eq:5.8}
\end{equation}
where at the last equality we have used not $M(\rho_0)$ but its inverse
function $\rho_0(M)$ which gives the density at the edge of the mass $M$
of higher density.  Now according to our cooling law higher densities
get denser faster, so the density ordering of fluid elements remains
the same.  If a given initial density $\rho_0$ evolves into a density $\rho$
after `time' $\tau$ then the function $\rho_0(M)$ will evolve into $\rho(M)$
with the {\bf same} $M$.  Thus the radius of the mass $M$ at time
$\tau$ will be given by analogy with (\ref{eq:5.8}) as 
\begin{equation}
\left.
\begin{array}{r}
(4/3)\pi r^3 \\
\pi R^2 \\
2\vert z \vert 
\end{array}
\right\}
= \int^M_0 1/\rho(M^\prime) dM^\prime\ .
\label{eq:5.9}
\end{equation}
Now near any smooth quadratic maximum the initial density takes the
form 
\begin{equation}
\rho_0 = \rho_m /(1+r^2/a^2) \equiv \rho_m (1+\alpha r^2/a^2)^{-1/\alpha} \ ,
\label{eq:5.10}
\end{equation}
[where $r^2$ should be replaced by $R^2$ or $z^2$ for the $2D$ or $1D$
cases]. 

Hence
\begin{displaymath}
M=(4/3)\pi r^3 \rho_m \left [1+O(r^2/a^2)\right ] \ ,
\end{displaymath}
so eliminating $r$ in favour of $M$
\begin{equation}
\rho_0(M) =
\left\{
\begin{array}{ll}
\rho_m \left \{1+\alpha [3M/(4\pi a^3 \rho_m)]^{2/3}\right \}^{-1/\alpha} 
	& {\rm spherical}\\
\rho_m \left \{1+\alpha M/(\pi a^2 \rho_m)\right \}^{-1/\alpha}
	& {\rm cylindrical}\\
\rho_m \left \{1+\alpha [M/(2a\rho_m)]^2\right \}^{-1/\alpha} 
	& {\rm planar}
\end{array}
\right.
\label{eq:5.11}
\end{equation}
Inserting this into  (\ref{eq:5.7}) we find
\begin{equation}
\rho_0 (M) = 
\left\{
\begin{array}{l}
\rho_c (m^{2/3}\pm 1)^{1/\alpha} \\
\rho_c (m      \pm 1)^{1/\alpha} \\
\rho_c (m^2    \pm 1)^{1/\alpha}
\end{array}
\right\}
= \rho_c(\tau)\rho_\ast (m)
\label{eq:5.12}
\end{equation}
where
\begin{equation}
m =  
\left\{
\begin{array}{ll}
3\alpha^{3/2} M/(4\pi a^3 \rho_m \vert 1-\tau/\tau_m \vert^{3/2}) 
	& {\rm spherical} \\
\alpha M/(\pi a^2 \rho_m \vert 1-\tau/\tau_m \vert ) 
	& {\rm cylindrical}\\
\alpha^{1/2}M/(2a \rho_m \vert 1-\tau/\tau_m \vert^{1/2}) 
	& {\rm planar}
\end{array}
\right.
\label{eq:5.13}
\end{equation}
and 
\begin{equation}
\rho_\ast = 
\left \{
\begin{array}{ll}
(m^{2/3}\pm 1)^{-1/\alpha} & {\rm spherical}\\
(m      \pm 1)^{-1/\alpha} & {\rm cylindrical}\\
(m^{2}  \pm 1)^{-1/\alpha} & {\rm planar}
\end{array}
\right.
\label{eq:5.14}
\end{equation}
To find $r$ we insert (\ref{eq:5.12}) into  (\ref{eq:5.9}) and deduce
\begin{equation}
(4/3)\pi r^3 \rho_c = \int^M_{0,M_{c}}(m^{2/3}\pm 1)^{1/\alpha}dM
\label{eq:5.15}
\end{equation}
where the lower limit is zero before collapse and the collapsed mass
$M_c$ afterward, as that collapsed mass does not contribute to the
radius.  Using the dimensional pieces of (\ref{eq:5.13}) and (\ref{eq:5.6})
to define a characteristic radius $r_c(t)$ we find $r=r_c(t)r_\ast (m)$ where
\begin{equation}
r^3_\ast = \int^m_{0,1} (m^{2/3}\pm 1)^{1/\alpha}dm\ ,
\label{eq:5.16}
\end{equation}
similarly
\begin{displaymath}
	R^2_\ast =
\left \{
		\begin{array}{ll}
		{{\alpha}\over{\alpha +1}} \left [(m+1)^{1+1/\alpha}-1 \right ]
			& \tau \geq \tau_m\\
		{{\alpha}\over{\alpha +1}} (m-1)^{1+1/\alpha} 
			& \tau \leq \tau_m\\
		\end{array}\\
\right.\\
\end{displaymath}
\begin{displaymath}
\begin{array}{ll}
	z_\ast = & \int^m_{0,1}(m^2\pm 1)^{1/\alpha}dm\\
\end{array}
\end{displaymath}
where $R_\ast = R/R_c(t), z_\ast = \vert z\vert/z_c$ and
\begin{equation}
\begin{array}{rcl}
r_c(\tau) & = & a \vert 1-\tau/\tau_m\vert^{1/2 + 1/3\alpha} 
		\alpha^{-1/2}(p_0/p)^{1/5}\\
R_c(\tau) & = & a \vert 1-\tau/\tau_m\vert^{1/2 + 1/2\alpha}
		\alpha^{-1/2}(p_0/p)^{3/10}\\
z_c(\tau) & = & a \vert 1-\tau/\tau_m\vert^{1/2 + 1/\alpha}
		\alpha^{-1/2}(p_0/p)^{3/5}
\end{array}
\label{eq:5.17}
\end{equation}
at least when $p$ is constant.

$z_c(\tau)<R_c(\tau)<r_c(\tau)$ as $\tau$ approaches $\tau_m$, and
(\ref{eq:5.17}) always holds near $\tau = \tau_m$ since $p(\tau)$ is not
sensitive to $\tau_m$.  Thus the planar collapse solution collapses
fastest, just like the gravitational case.  This strongly suggests
that the spherical solutions are unstable to a flattening instability
and we shall see that this is indeed true.

Equations (\ref{eq:5.14}) and (\ref{eq:5.16}) together constitute
parametric equations for the density profile $\rho_\ast(r_\ast)$ in
terms of the parameter $m$.  From (\ref{eq:5.12}) it is already clear
that the solution is self-similar both before and after the moment
when the density first becomes singular.  The spherical solutions with $\alpha = 1$ are plotted in Figure 3.   This self-similarity rests
on the approximate form of $\rho_0$ close to its maximum $\rho_m$, so
it only holds for times close to that moment.  One may find
$\rho_\ast(r_\ast)$, or rather its inverse function
$r_\ast(\rho_\ast)$, explicitly for the special cases when $\alpha =
2/n$, with $n$ an integer, but we keep to the general case and look at
the asymptotic forms for small and for large $r_\ast$.

\noindent
{\bf Pre-collapse}
\begin{equation}
\left.
\begin{array}{cc}
	{\rm For}\ r_\ast \ {\rm small}\ r^3_\ast = m+(3/5)\alpha^{-1}
	  m^{5/3}
		& {\rm so}\ \\
	m=r^3_\ast -(3/5)\alpha^{-1} r^5_\ast 
		& {\rm and\ from}\  (\ref{eq:5.14}) \\
	\rho_\ast \simeq (1+r^2_\ast)^{-1/\alpha} 
		&  r_\ast \ {\rm small}
\end{array}
\right \}
\label{eq:5.18}
\end{equation}
\begin{equation}
\left.
\begin{array}{cc}
{\rm for}\ r_\ast \gg 1 & r^3_\ast = {{3\alpha }\over{3\alpha+2}}
m^{2/(3\alpha)}
\left [m + {{2+3\alpha}\over{2+\alpha}} m^{1/3} \ldots \right ]\\
& m=\eta^{-\eta} r^{3\eta}_\ast \left [1-{{3\eta}\over{3+2\eta}}
\alpha^{-1} \eta^{1+2\eta/3} r_\ast^{-2\eta} + O(r^{-4\eta}_\ast)\right ]\\ 
{\rm where} & 
 \eta = 3\alpha/(3\alpha +2) \\
{\rm so\ for\ large\ }r_\ast,\ (\ref{eq:5.14})\ {\rm gives} & 
\rho_\ast = (1+\eta^{-2\eta/3} r^{2\eta}_\ast)^{-1/\alpha}\propto r^{-6/(3\alpha +
2)}_\ast \\
\end{array}
\right \}
\label{eq:5.19}
\end{equation}
Thus at the moment when the central density becomes singular we find a
power law density profile with index $-6/(3\alpha+2)$.  That is
$-12/13$, $-6/5$ and $-12/7$ for $\alpha = 3/2,\ 1$ and $1/2$
respectively.

\noindent
The $r^{-12/7}$ profile is the same as that for Penston's cold
free-fall solution for a self gravitating spherical cloud, \citep{pen69}.

\noindent
{\bf Post collapse}
\begin{equation}
\begin{array}{c}
{\rm For\ times\ later\ than}\ \tau_m\ {\rm the\ central\ point\
mass}\ M_c \ {\rm grows}\\
M_c = \left [(\tau/\tau_m)-1\right ]^{3/2}(4/3)\pi a^3 \rho_m
\alpha^{-3/2}
\end{array}
\label{eq:5.20}
\end{equation}
which comes directly from expression (\ref{eq:5.13}) with $m=1$.  Of
course (\ref{eq:5.20}) only holds when $\tau$ is close to $\tau_m$ when
approximation (\ref{eq:5.10}) is valid.  However, it is easy enough to
determine $M_c(\tau)$ directly from the initial conditions.  Let
$M[\rho_0]$ be the mass within the $\rho_0$ density contour in the initial
conditions and suppose it all has density $\geq \rho_0$ for all $\rho_0$
above some minimum.  Then from our expression for $\tau_c$ above
(\ref{eq:5.5}) the mass that has collapsed by time $\tau$ is the mass
that had initial density greater than $\left [(2/5)\alpha K
p^{1-\alpha}_0 \tau \right ]^{-1/\alpha}$ i.e., 
\begin{equation}
M_c = M\left [\left ((2/5)
\alpha K p^{1-\alpha}_0 \tau\right )^{-1/\alpha}\right ]
\label{eq:5.21}
\end{equation}
Setting $m=1+\tilde{m}$ in (\ref{eq:5.16}), the post collapse similarity
solution for times a little greater than $\tau_m$ is (see Figure 3c)
\begin{equation}
\begin{array}{lllll}
	& r^3_\ast & = & (2/3)^{1/\alpha}\left
	[\alpha/(\alpha+1)\right ] \tilde{m}^{1+1/\alpha} + \ldots & \\
	& \rho^{-\alpha}_\ast & = & (2/3)\tilde{m} + \ldots &\\
{\rm hence} 
	& \rho_\ast & = & 
		\left \{(3/2)\left [\alpha/(\alpha+1)\right
	]r^{-3}_\ast \right \}^{1/(\alpha +1)} + \ldots & 
			{\rm small}\ r_\ast
\end{array}
\label{eq:5.22}
\end{equation}
i.e., a power law of index $-6/5,\ -3/2,\ {\rm or}\ -2\ {\rm for}\
\alpha = 3/2,\ 1,\ 1/2$.  At $r>r_c(t)$ these power laws change over
to the power $-6/(3\alpha+2)$ derived earlier.
\section{Exactly Self-Similar Solutions}
We derived these solutions as approximations, only good for times
close to $\tau_m$ for regions not too far from the maximum density.
However if, instead of postulating a general initial profile and then
approximating it near maximum by (\ref{eq:5.10}), we had taken an initial
profile defined by (\ref{eq:5.16}) and (\ref{eq:5.14}) then our solution
would be exactly self-similar at all times and all positions.  Perhaps
the simplest of these is that given by Newton's law of cooling
proportional to temperature.  Then (\ref{eq:5.16}) integrates and yields
for $\alpha = 1$
\begin{displaymath}
\begin{array}{lllc}
r^3_\ast 	& = & m+(3/5)m^{5/3} 		& \tau < \tau_m \\
		&   & (3/5)(m^{5/3}-1)-m+1 	& \tau > \tau_m \\
\multicolumn{3}{c}{{\rm while\ (\ref{eq:5.14}) \ yields\ }}
		& \rho_\ast = 1/(m^{2/3}\pm 1) 
\end{array}
\end{displaymath}
Asymptotically $\rho_\ast \rightarrow r^{-6/5}_\ast $ for all large
$r_\ast$ so to keep the pressure constant $T \rightarrow
r_\ast^{6/5}$.  

Similarly the `exact' solution for the cylindrical
case gives $\rho$ explicitly
\begin{displaymath}
\rho = \rho_c(t)/\left [U+(1+\alpha^{-1})R^2_\ast \right ]^{1/(\alpha+1)} 
\end{displaymath}
where $U=1$ for $\tau<\tau_m$ and $0$ otherwise; $\rho_c(\tau)$ is
still given by (\ref{eq:5.6}) in all three geometries.

For $\alpha =1$ we have for the planar case
\begin{displaymath}
\begin{array}{lll}
		& \rho_\ast = 1/(m^2\pm 1)  \\
{\rm and}	& z_\ast = \vert z\vert/z_c = & \left\{
\begin{array}{ll}
(1/3)m^3 + m 	& \tau < \tau_m \\
(1/3)m^3-m+2/3 	& \tau > \tau_m
\end{array}
\right.
\end{array}
\end{displaymath}

We have obtained all these solutions under the assumption that the
cooling is so slow that we may take the pressure to be almost constant
- i.e., the system evolves quasi-statically through a sequence of
equilibria.  Nevertheless if it evolves in finite time there must be
small motions and very small accelerations to drive these small
motions.  To find these motions and the small pressure gradients that
drive them we use the equations of fluid mechanics.  By mass
conservation
\begin{displaymath}
D{\rm ln} \rho/Dt = - {\rm div}\ {\bf u} = - r^{-2}\partial/\partial r(r^2u)
\end{displaymath}
where $u$ is the radial component of ${\bf u}$.  We know $D{\rm ln}
\rho/Dt$ from (\ref{eq:5.1}) so using (\ref{eq:5.16})
\begin{displaymath}
u= -(1/5)(\dot{p}/p)r-(2/15)Kp^{1-\alpha} \rho^\alpha_c r_c  
	r^{-2}_\ast \int^m_{0,1} (m^{2/3}\pm 1)^{(1-\alpha)/\alpha}
	dm\ .
\end{displaymath}

When $\alpha =1$ the final integral simplifies to $m$ for $\tau<0$ and
$m-1$ for $\tau>0$.  Such solutions are plotted for $\dot{p}=0$ as
Figure 4.

In $2D$
\begin{displaymath}
u = - (3/10) \left ({{\dot{p}}/{p}}\right )r -(1/5)Kp^{1-\alpha}
\rho^\alpha_c R_c r^{-1}_\ast \alpha
\left\{
\begin{array}{ll}
(m+1)^{1/\alpha} -1;\ \tau<0\\
(m-1)^{1/\alpha} ; \ \tau>0 \ .
\end{array}
\right.
\end{displaymath}

In the planar case with $\alpha =1$
\begin{displaymath}
u = -(3/5)(\dot{p}/p)z - (2/5)Kp^{1-\alpha}\rho^{\alpha}_c
	r_c m\ .
\end{displaymath}

Notice that similarity only extends to the velocity fields when either
$\dot{p}$ is zero or $\dot{p}\propto p^{2-\alpha}\rho^\alpha_c$.  From
the above $Du/Dt$,  and thence the
small pressure gradient can be calculated.  Spherical self-similar
solutions for the outer parts of cooling flows have been found
numerically by \citet{ber89}.  His solutions are valid through the
sonic region.
\section{Flattening into Sheets}

We saw in section 5 under (\ref{eq:5.17}) that the collapse time
depended on the dimension and the planar collapse was fastest. We were
thus led to believe that density maxima would flatten and the cooling
instabilities would lead to curtain-like sheets of high density.  To
check that this is indeed the case, Marcus
Br\"{u}ggen, ran some 3D and 2D simulations.  In these the
full equations of fluid dynamics were integrated and the cooling was
switched on over a few sound-crossing times to avoid transients.
While the pressure was initially constant in space it was allowed to
vary in space and time.  While the cooling was slow it was not so slow
that pressure remained constant and indeed it fell drastically in the
cooling sheets that developed.  Figures 5a and 5b which 
contour the density distribution initially, and after cooling, show
clearly the development of flat sheets of high density just like those
studied in planar geometry by \citet{bur00}.  The high density and low
sound speed in these sheets make them ideal for gravitational
instability and fragmentation into clusters of stars.
\section{Conclusions}
While some sort of census of the contents of the Universe is now
possible, the true nature of the vast majority of it is unknown.
Astronomy is still young.

Many more halo white dwarfs are now being found and this is likely to
change our concept of what the galaxy is like.  They probably make a
significant contribution to the dark matter (5\% - 10\%), however
there remain serious doubts on the reality of this interpretation.
This is an exciting and controversial subject that deserves our attention.

Pristine white dwarf configurations exist for objects that never burn
hydrogen but no-one has devised a way of getting bodies to that state
so the large numbers of halo white dwarfs now being found are probably
normal burnt out stellar remnants.  

Cooling instabilities may be the method by which density enhancements
start in many different areas of astronomy; gravity may be only
important in the final stages of star formation.

\section{Acknowledgements}
Firstly to the American Astronomical Society for the honour of asking
L-B to give a Russell Lecture, the first of a new Millenium.  Secondly
to those who have been especially helpful; Marcus
Br\"{u}ggen using his expertise in simulating cooling clouds,
M. J. Irwin, R. F. Griffin \& F. van Leeuwen for their observations;
and O. Lahav for Cosmological consultations.

Finally to Astronomy for the great fun it gives us all.

\anchor{ftp://www.aas.org/pubs/}{AAS ftp site}.
For technical support, please write to

\email{aastex-help@aas.org}.

\appendix

\clearpage
\begin{figure}
\plotone{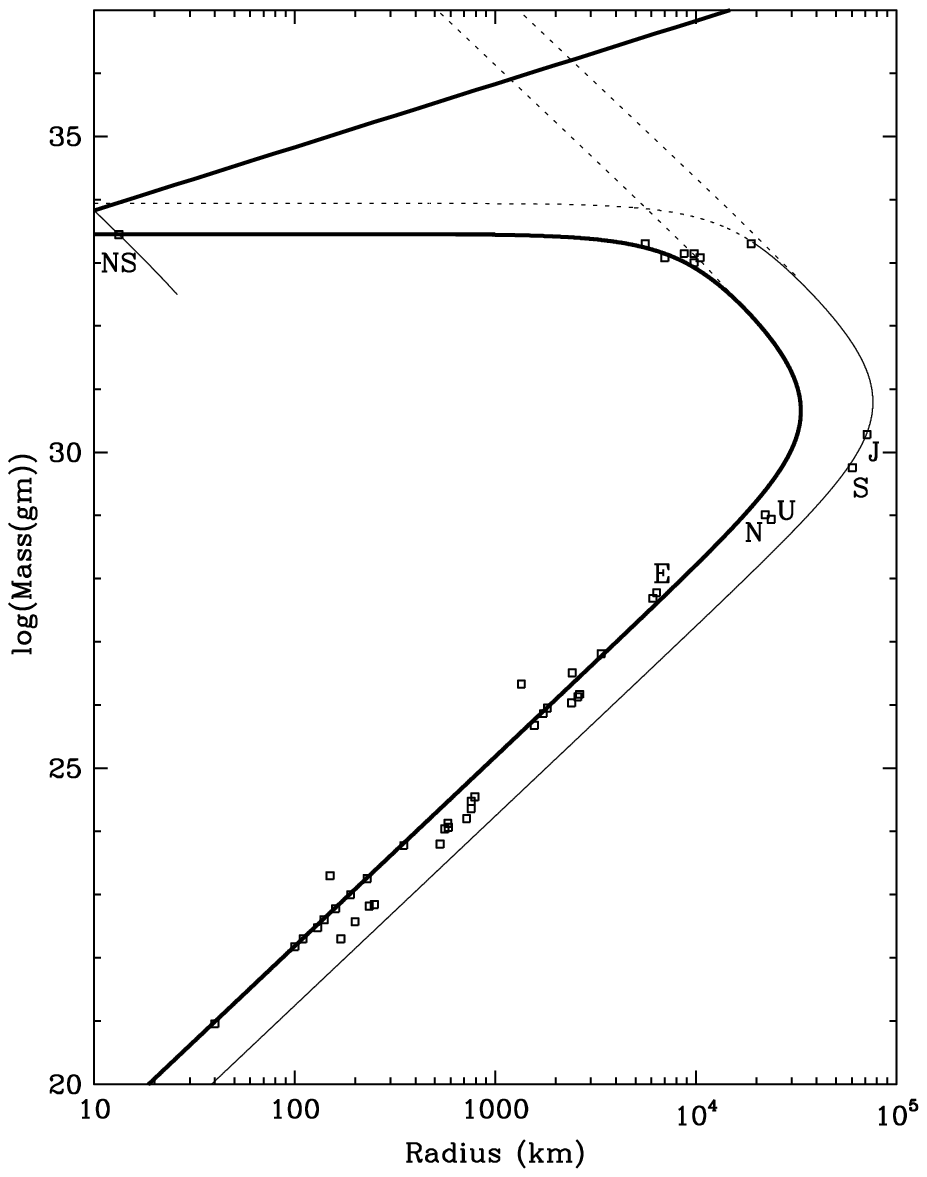}
\caption {Figure 1. The heavy line on the left gives the mass-radius relation
for cold bodies with $\rho_0 = 3.65\ {\rm g\ cm}^{-3}$ and $\mu = 2$,
suitable for rocky planets and normal white dwarfs.  The heavy line
above gives the black hole radius.  The lighter line is for a mixture
with 75\% Hydrogen and 25\% Helium, $\mu = 8/7$ suitable for gas-rich
planets, brown dwarfs and pristine white dwarfs.  This sequence ends
at $1.1M_\odot$ due to pressure induced, (pycno-nuclear) reactions.  Stable
bodies could exist between 0.2 and 1.1$M_\odot$ but no plausible
formation path has been found.  The dashed lines extrapolate the
non-relativistic formula for degeneracy into the region of its
invalidity.  The neutron star line is parallel to these but displaced
from the pure hydrogen line by the factor $(m_e/m_n)$ in radius.}
\end{figure}

\begin{figure}
\plotone{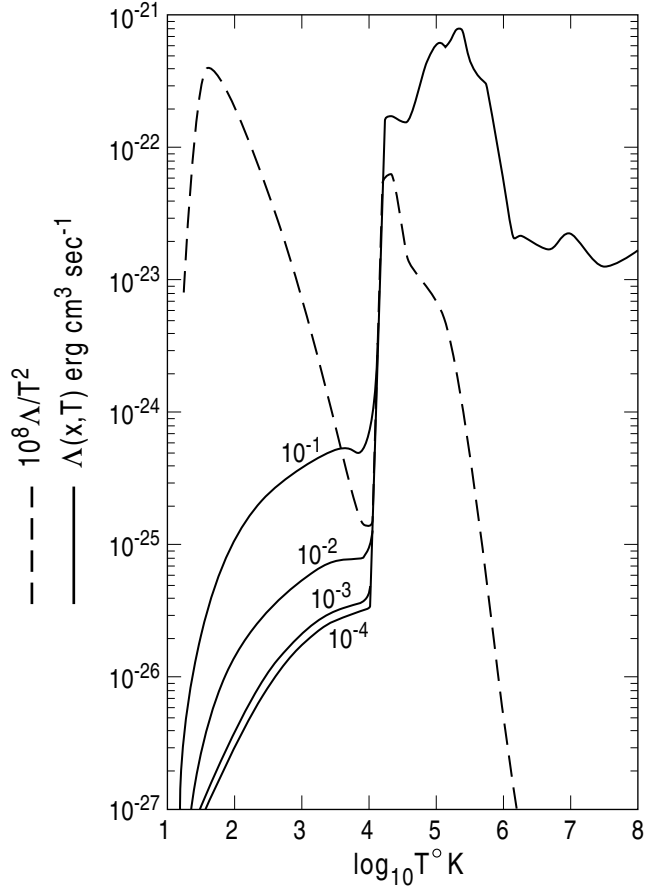}
\caption{Figure 2.  The cooling function $\Lambda(T)$ of Dalgarno \&McCray
 with superimposed (dashed) the graph of
$10^8\Lambda(T)/T^2$.  There is cooling instability causing
condensations in an otherwise homogeneous region wherever the latter
graph is falling, i.e., everywhere with $T>30K$ except the region near
$10^4K$ where hydrogen is  partially ionised.  Cooling instability
should be prevalent throughout astronomy.}
\label{fig2}
\end{figure}

\begin{figure}
\plotone{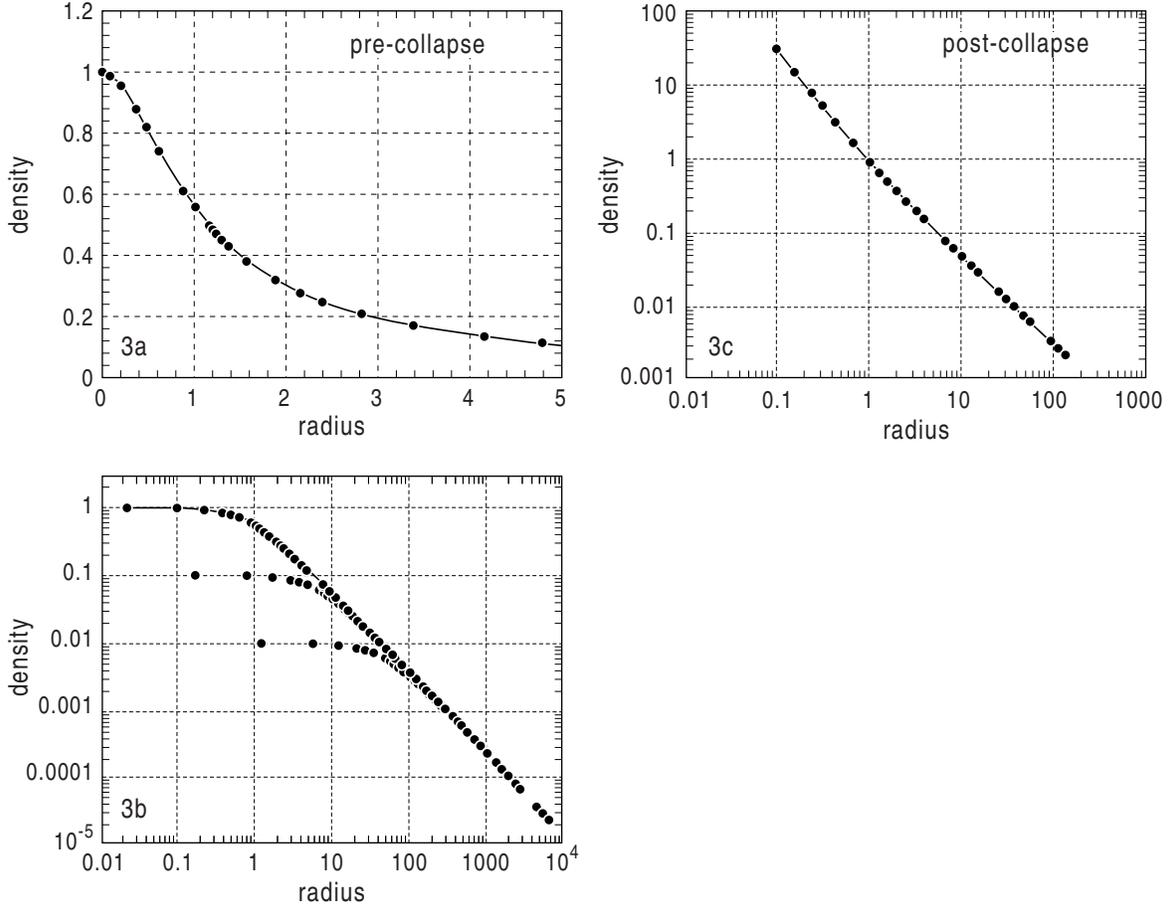}
\caption{{\bf Figure 3a}. The density profile for the spherical self-similar
solution with $\alpha=1$. 
 {\bf Figure 3b}.  The log density -- log radius profile at times
${{\tau}\over{\tau_m}}= -1, -10, -100$ showing how the scale of the
profile shrinks as $\tau=0$ is approached.
{\bf Figure 3c}.  The same as 3b but after the centre collapses.  At
$\tau=1$ there is a growing central mass and the radius at which the
power law changes from $-3/2$ to $-6/5$ increases like
$(\tau/\tau_m-1)^{5/6}$ again for $\alpha = 1$.}
\end{figure}

\begin{figure}
\plotone{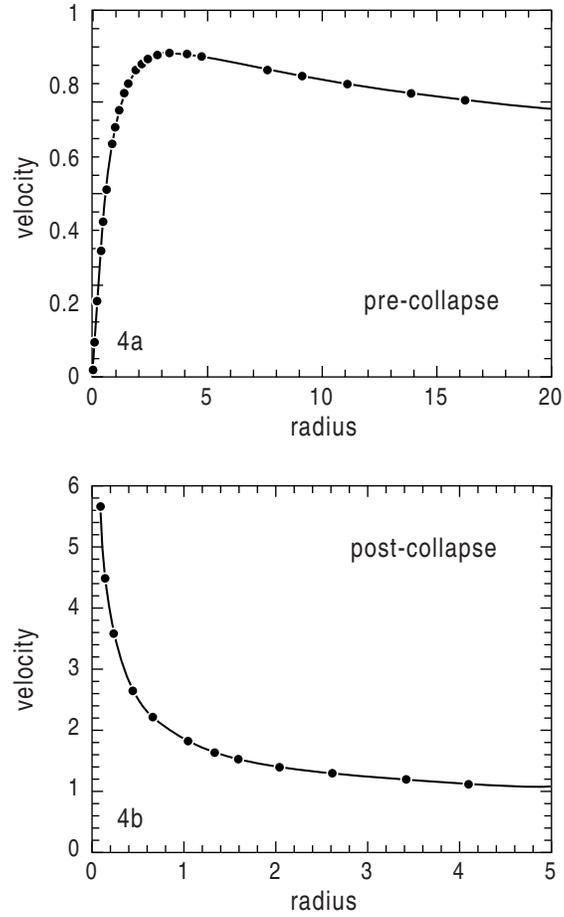}
\caption{
{\bf Figure 4a}. The inward radial velocity profile prior to collapse of
the core.
{\bf Figure 4b}.  The same after the point mass develops $u\propto r^{-1/2}$
for small $r$ and $u\propto r^{-1/5}$ for large $r$.  The transition
point moves outward. }
\label{fig4}
\end{figure}

\begin{figure}
\plotone{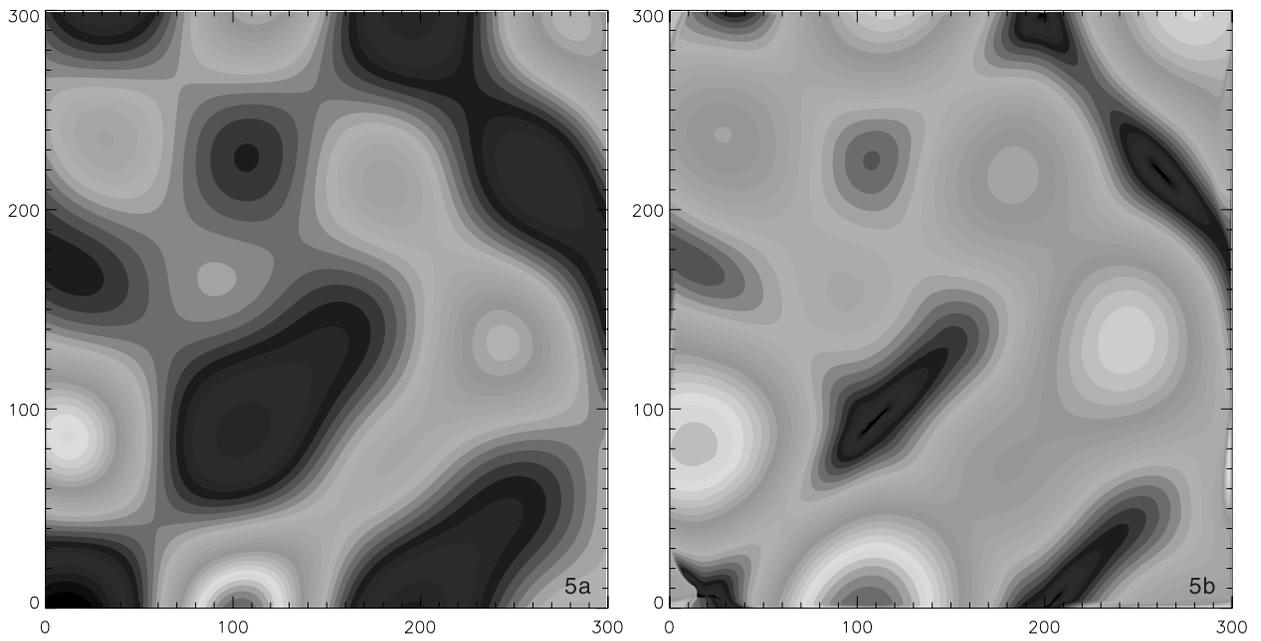}
\caption{{\bf Figure 5a}.  Initial density contours of M. Bruggen's
simulation of cooling flow. {\bf Figure 5b}.  Density contours after
thin sheets develop which collect the mass.  }
\label{fig5a}
\end{figure}

\end{document}